\documentclass[%
  aip,
  amsmath,amssymb,
  floatfix,%
  reprint,%
]{revtex4-1}

\usepackage{graphicx}
\usepackage{dcolumn}
\usepackage{bm}
\usepackage[utf8]{inputenc}
\usepackage[T1]{fontenc}
\usepackage{mathptmx}
\usepackage{hyperref}
\usepackage{booktabs}
\newcommand{\SiGe}{Si$_{1-x}$Ge$_x$}
\newcommand{\DV}{\Delta E_v}
\newcommand{\DC}{\Delta E_c}
\newcommand{\VIF}{\Delta V_{\mathrm{IF}}}
\newcommand{\Vbar}{\overline{V}}

\begin{document}

\title{First-principles predictions of band alignment in strained Si/\SiGe\ and Ge/\SiGe\ heterostructures}

\newcommand{\McGillAffil}{Department of Physics, McGill University, Montr\'eal, QC H3A 0G4, Canada}
\newcommand{\NanoAffil}{Nanoacademic Technologies Inc., Montr\'eal, Suite 802, 666 rue Sherbrooke Ouest, Montréal, Québec, Canada H3A 1E7}

\author{Nathaniel M.\ Vegh}
\email{nathaniel.vegh@mail.mcgill.ca}
\affiliation{\NanoAffil}
\affiliation{\McGillAffil}

\author{Pericles Philippopoulos}
\affiliation{\NanoAffil}

\author{Rapha\"el J.\ Prentki}
\affiliation{\NanoAffil}

\author{Wanting Zhang}
\affiliation{\NanoAffil}

\author{Yu Zhu}
\affiliation{\NanoAffil}

\author{F\'elix Beaudoin}
\affiliation{\NanoAffil}

\author{Hong Guo}
\affiliation{\McGillAffil}

\begin{abstract}
Accurate band offsets are essential for predictive continuum modeling of nanostructures such as quantum wells and quantum dots formed in strained Si/\SiGe\ and Ge/\SiGe\ heterostructures. Experimental offset data for these systems remain sparse away from endpoint compositions, which makes composition-dependent design difficult. We use atomistic, first-principles density functional theory to compute valence- and conduction-band offsets across the full range $0 \le x \le 1$. Random alloying is treated with special quasirandom structures, interface lineup terms are extracted from macroscopically averaged local Kohn--Sham potentials in thick periodic superlattices, valence-band spin--orbit coupling is included through species-resolved Mulliken weights, and conduction-band edges are refined using the screened hybrid Heyd--Scuseria--Ernzerhof functional. The resulting offsets show pronounced composition nonlinearity beyond the linear models explored in previous works, agree with experimental benchmarks, and reproduce the high-Ge slope change in the relaxed-alloy band gap. Analytic fitting expressions are provided for direct use in simulations, facilitating practical design of modern quantum technology devices.
\end{abstract}

\maketitle


Strained Si/\SiGe\ and Ge/\SiGe\ heterostructures are central to modern electronic and quantum devices, including high-mobility transistors, modulation-doped quantum wells, and electron and hole quantum dots.\cite{Schaffler1997,Paul2004,Yang2004,Zwanenburg2013,Vandersypen2017,Scappucci2021}
Recent Si/\SiGe\ spin-qubit experiments have demonstrated high-fidelity two-qubit operation and multiqubit control,\cite{Noiri2022,Xue2022,Mills2022,Philips2022} while strained-Ge/\SiGe\ quantum wells provide a complementary hole-spin platform with strong spin-orbit coupling and no valley degeneracy, with experiments now scaling to ten- and eighteen-qubit germanium arrays.\cite{Scappucci2021,John2025,Dijkema2026}
In these platforms, valence- and conduction-band offsets enter directly as material parameters in continuum electrostatics workflows used to predict confinement potentials, tunnel barriers, and qubit performance.\cite{Yang2004,Andersen2009,Beaudoin2022}
Reliable offsets across composition are therefore required to separate genuine device-physics from uncertainty in material inputs.

A first-principles description of band alignment decomposes the offset into (i) a bulk band-edge term, defined relative to a reference potential in each material, and (ii) an interface-lineup term, which captures charge rearrangement and the resulting interface dipole.\cite{VandeWalleMartin1987,Dandrea1992,VandeWalle1999,DiLiberto2021}
Compared with Anderson's electron-affinity rule,\cite{Anderson1962} which aligns materials using vacuum-referenced electron affinities, this decomposition uses an explicit interface electrostatic lineup and avoids the ambiguities of vacuum-level alignment.\cite{VandeWalleMartin1987,DiLiberto2021}
For isovalent Si/Ge-derived interfaces, prior first-principles analyses indicate that interface dipoles from interdiffusion or atom exchange across the interface are typically small, while strain-driven band-edge shifts and splittings remain central to quantitative offsets.\cite{VandeWalleMartin1987,Dandrea1992,VandeWalle1999}
In strained SiGe systems, the lineup problem is coupled to strain-induced band-edge shifts and splittings,\cite{VandeWalle1999,RiegerVogl1993,Colombo1991} while random-alloy disorder introduces finite-size and configuration-sampling effects that must be quantified.

Experimentally, band-offset extraction in Si/Ge/SiGe systems remains challenging and is often limited to specific structures or compositions.
This is particularly relevant because the most common spin-qubit regimes already cluster around strained-Si wells on Si-rich SiGe barriers with $x\approx 0.25$--0.33 and strained-Ge wells on Ge-rich SiGe barriers with $x\approx 0.6$--0.9, often near $x\approx 0.8$.\cite{Schaffler1997,Zwanenburg2013,Scappucci2021}
Representative benchmarks include core-level photoemission on pseudomorphic Si/Ge interfaces,\cite{Schwartz1989} Ge/\SiGe\ photoreflectance at a few Ge-rich barrier compositions,\cite{Yaguchi1994} selected Si/\SiGe\ measurements compiled by Van de Walle,\cite{VandeWalle1999} and MOS-capacitor extraction of the strained-Si/\SiGe\ conduction offset at $x=0.2$.\cite{Maiti2004}
These studies provide key anchor points, but they do not by themselves deliver a dense, internally consistent dataset for both strained Si/\SiGe\ and Ge/\SiGe\ across $0 \le x \le 1$.

On the modeling side, parameter compilations and analytic models are invaluable for simulation practice,\cite{Schaffler1997,Paul2004,Yang2004,VandeWalle1999,Virgilio2006} yet they are assembled from heterogeneous experiments and theoretical assumptions and are not always internally consistent across strain states, compositions, and interface structures.
This gap motivates a single, internally consistent first-principles dataset across the full composition range.
Here we apply one procedure throughout: explicit interface-lineup extraction in thick periodic superlattices, special quasirandom structure (SQS) treatment of random alloys, a Mulliken-weight spin--orbit-coupling (SOC) correction of the valence edge, and hybrid-functional refinement of conduction edges.

All calculations are performed using the real-space density-functional theory (DFT) package RESCU,\cite{MichaudRioux2016,Nanoacademic}
and employ the Perdew--Burke--Ernzerhof (PBE) generalized-gradient approximation (GGA) for the exchange--correlation (XC) energy.\cite{PBE1996}
To correct the semilocal band-gap error in the conduction offsets, we later refine the conduction-edge terms with the Heyd--Scuseria--Ernzerhof HSE06 hybrid functional\cite{Heyd2003,Heyd2006} while retaining the PBE lineup workflow.
HSE06 calculations use the standard screened hybrid form with 25\% short-range exact exchange and a screening parameter corresponding to approximately 0.2~\AA$^{-1}$.\cite{Heyd2003,Heyd2006}
The wavefunctions are represented in a linear combination of atomic orbitals (LCAO) basis.\cite{MichaudRioux2016}
Since RESCU is a real-space LCAO code, there is no plane-wave kinetic-energy cutoff; the analogous numerical parameter is the real-space-grid spacing, which is specified as an upper bound so that the grid actually used is at least as fine. The grid-spacing bounds are 0.14~\AA\ for the bulk alloy cells, 0.20~\AA\ for the strained-Si bulk reference cells, and 0.18~\AA\ for the heterostructure superlattices.
Bulk band-edge calculations are performed in 64-atom supercells (\SiGe\ SQS built from a $2\times 2\times 2$ replication of the conventional cubic cell)\cite{Zunger1990} using a $3\times 3\times 3$ Monkhorst--Pack $k$-point grid.\cite{MonkhorstPack1976}
For lineup extraction, we use thick periodic heterostructure superlattices (512 atoms) sampled with a $3\times 3\times 1$ Monkhorst--Pack $k$-point grid.
All bulk and heterostructure calculations use three-dimensional periodic boundary conditions defined by the supercell lattice vectors and sampled over the corresponding Brillouin zone.
Structural relaxations are performed at fixed in-plane lattice constants set by the relaxed alloy buffer (epitaxial constraint), while internal coordinates and the out-of-plane lattice parameter are relaxed to the numerical force/stress convergence thresholds.
In strained bulk reference cells, $E_v$ and $E_c$ denote the valence-band maximum (VBM) and conduction-band minimum (CBM) energies and are taken as the highest occupied and lowest unoccupied Kohn--Sham eigenvalues over the sampled Brillouin-zone mesh.\cite{VandeWalleMartin1987,Dandrea1992,DiLiberto2021}
Operationally, $E_v$ ($E_c$) is the maximum (minimum) eigenvalue with occupation threshold $n_{\mathrm{occ}}>10^{-3}$ ($<10^{-3}$), and bulk band edges are measured relative to the cell-averaged local potential $\langle v_{\mathrm{dh}}+v_{\mathrm{na}}+v_{\mathrm{xc}}\rangle$, where $v_{\mathrm{dh}}$ is the Hartree potential generated by the self-consistent charge redistribution, $v_{\mathrm{na}}$ is the neutral-atom potential used to assemble the atomic lattice, and $v_{\mathrm{xc}}$ is the exchange--correlation term.\cite{MichaudRioux2016,DiLiberto2021}

\begin{figure}[t]
\centering
\includegraphics[width=\columnwidth]{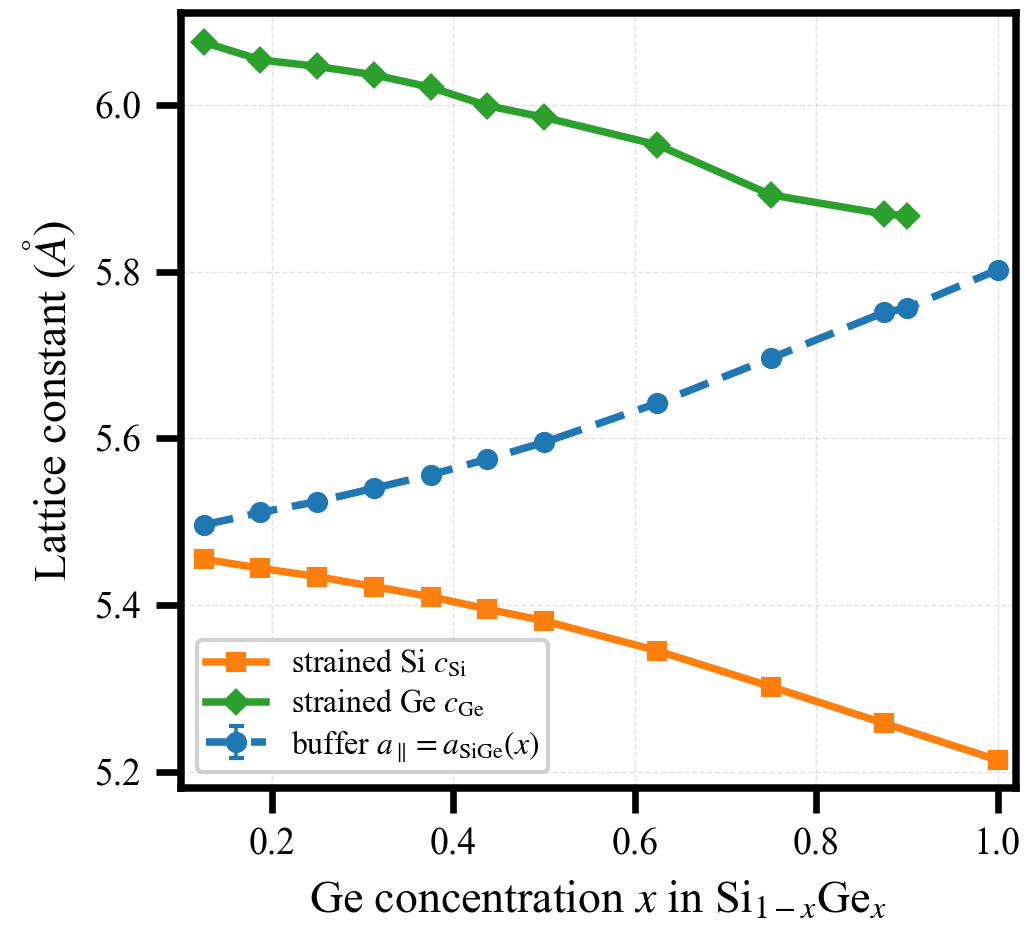}
\caption{\label{fig:epitaxy}
Epitaxially constrained lattice constants used in this work.
The relaxed-alloy buffer lattice constant $a_{\parallel}=a_{\mathrm{SiGe}}(x)$ is compared with the relaxed out-of-plane lattice constants $c_{\mathrm{Si}}$ and $c_{\mathrm{Ge}}$.
Error bars on $a_{\parallel}$ show a representative $\pm 0.001$~\AA\ uncertainty from EOS-fitted SQS minima.
The strained-Ge series is shown up to $x=0.9$; $x=1$ is the Ge/Ge homoepitaxial limit.}
\end{figure}

\begin{figure}[t]
\centering
\includegraphics[width=0.92\columnwidth]{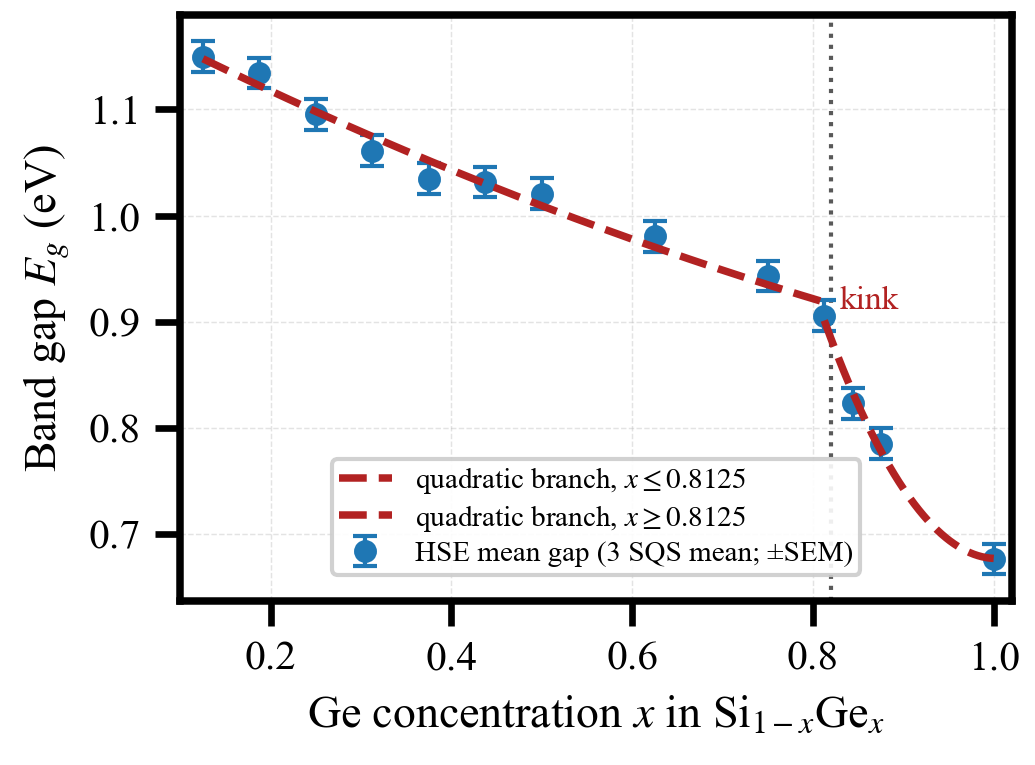}
\caption{\label{fig:gap_relaxed}
HSE band gap of relaxed \SiGe\ across composition.
Points show three-SQS means with representative standard-error bars derived from a characteristic full spread of about 50~meV.
Dashed guides are overlapping quadratic fits with a crossover near $x\approx 0.82$.}
\end{figure}

We consider strained Si and strained Ge on a relaxed \SiGe\ buffer at composition $x$, each biaxially constrained in plane to the buffer lattice constant.
The buffer lattice constant $a_{\parallel}=a_{\mathrm{SiGe}}(x)$ is obtained by equation-of-state (EOS) total-energy minimization of isotropically scaled \SiGe\ SQS cells.
Figure~\ref{fig:epitaxy} summarizes the resulting lattice constants.
For the random alloy, we construct SQS supercells designed to reproduce the pair correlations of a random SiGe alloy within a finite supercell.\cite{Zunger1990}
For the primary interface-offset dataset reported in Supplementary Table~S1, one SQS superlattice is used at each composition. For Fig.~\ref{fig:gap_relaxed}, three SQS configurations are averaged per composition. Additional repeated-SQS tests for the bulk alloys, together with repeated alloy-configuration tests and numerical convergence checks ($k$-mesh and grid resolution) for the interface cells, were used to quantify how sensitive the computed offsets are to the alloy configuration and the numerical settings.
The relaxed-alloy HSE gap shows a representative three-SQS full spread of about 50~meV, while the extracted lineup is reproducible at about the 10~meV level across repeated configuration and numerical tests; these values set the representative uncertainty model used for the offset error bars.
We did not use phonon spectra as an additional stability criterion because SiGe random alloys and coherent Si/\SiGe\ and Ge/\SiGe\ heterostructures are experimentally established materials; we focus on modeling the relaxed epitaxial strain state rather than predicting a new crystalline phase.

With the strain state and bulk reference quantities defined, we adopt the potential-lineup decomposition introduced by Van de Walle and Martin.\cite{VandeWalleMartin1987,Dandrea1992,VandeWalle1999,DiLiberto2021}
For a heterojunction between materials $A$ and $B$, the valence-band offset (VBO) is
\begin{equation}
\DV
=
\left[
E_{v}^{B}-\Vbar^{B}
\right]
-
\left[
E_{v}^{A}-\Vbar^{A}
\right]
+
\left[
\Vbar^{B}-\Vbar^{A}
\right]_{\mathrm{IF}}
+
\Delta E_{v,\mathrm{SOC}}^{B/A},
\label{eq:vbo}
\end{equation}
and similarly for the conduction-band offset (CBO),
\begin{equation}
\DC
=
\left[
E_{c}^{B}-\Vbar^{B}
\right]
-
\left[
E_{c}^{A}-\Vbar^{A}
\right]
+
\left[
\Vbar^{B}-\Vbar^{A}
\right]_{\mathrm{IF}}
+
\Delta E_{c}^{\mathrm{SOC}}.
\label{eq:cbo}
\end{equation}
To make the terms in Eqs.~(\ref{eq:vbo}) and (\ref{eq:cbo}) explicit, we define the bulk macroscopic-averaged potential and lineup terms as follows.
For each bulk reference cell $X\in\{A,B\}$, the bulk macroscopic-averaged potential is
\begin{equation}
\Vbar^{X}=\frac{1}{\Omega_X}\int_{\Omega_X}\left(v_{\mathrm{dh}}+v_{\mathrm{na}}+v_{\mathrm{xc}}\right)\,d^3r,
\label{eq:vbar_bulk}
\end{equation}
where $\Omega_X$ is the bulk-cell volume.
Using the same local-potential reference in both bulk and superlattice calculations ensures strict internal consistency across the workflow.
The interface lineup term is
\begin{equation}
\VIF \equiv \left[\Vbar^{B}-\Vbar^{A}\right]_{\mathrm{IF}},
\label{eq:vif_def}
\end{equation}
where the subscript IF denotes the interface contribution extracted from an explicit periodic $A/B$ superlattice and corresponds to the macroscopic potential step associated with the interface dipole.\cite{VandeWalleMartin1987,Dandrea1992}
In this work we apply SOC corrections only to the valence term and neglect an explicit $\Delta E_{c}^{\mathrm{SOC}}$, consistent with prior SiGe band-offset treatments in which SOC is needed to determine the valence-band maximum accurately, while conduction offsets are obtained from the relevant $X$- and $L$-valley positions and band-gap differences.\cite{VandeWalleMartin1987,VandeWalle1999}

The lineup in Eq.~(\ref{eq:vif_def}) is extracted by computing the self-consistent local Kohn--Sham potential in a periodic heterostructure superlattice, planar-averaging along the growth direction $z$, and smoothing to remove atomic-scale oscillations.\cite{VandeWalleMartin1987,Baldereschi1988}
All superlattices are oriented along the cubic [001] growth direction and constructed as coherent (001) Si/SiGe heterostructures with a $2\times 2$ in-plane repeat.
A representative lineup cell contains 8 conventional cells of strained Si and 8 conventional cells of \SiGe\ (SQS) along $z$ (256 atoms per side, 512 atoms total), yielding two interfaces per periodic supercell.
The lineup is taken as the difference between plateau values in bulk-like regions sufficiently far from the interfaces.
We emphasize that no vacuum region is used: the periodic superlattice contains only coherent A/B interfaces and avoids surface dipoles and termination sensitivity inherent to vacuum-level alignment.
Here, plateau selection refers to choosing the bulk-like averaging windows on each side of the interface from which the potential step is measured.
In practice, we planar-average the local potential along $z$, smooth it with a periodic error-function kernel of width 2.0~\AA\ applied twice, and evaluate the plateau values from fixed bulk-like windows within each slab, away from the interfacial oscillations.
By increasing the number of conventional cells per side from $N=2$ to $12$, the extracted $\VIF$ varies by only about 0.02~eV once $N \ge 8$, with similarly small changes observed in the other sampled compositions.
To further quantify the numerical and configurational robustness of the extracted lineup, we use a smaller test system: a Si/\SiGe\ ($x=0.5$) superlattice with a $1\times1$ in-plane repeat and $N=4$ conventional cells per side (64 atoms). This reduced cell is inexpensive enough for systematic one-at-a-time tests, and it also enables the direct PBE/HSE comparison presented below, which would be prohibitive in the 512-atom production cells. In this cell, refining the $k$ mesh from $3\times3\times1$ to $4\times4\times1$ changes $\VIF$ by $-16.2$~meV, refining the real-space-grid spacing bound from 0.20 to 0.18~\AA\ changes it by $+12.9$~meV, and repeating the calculation with three independent Si/Ge site arrangements in the alloy slab changes it by only 7.8~meV in total (sample standard deviation 4.0~meV).

We apply a valence-band SOC correction based on species-resolved Mulliken populations:\cite{Mulliken1955}
\begin{equation}
\Delta E_{v}^{\mathrm{SOC}}(\mathrm{SiGe})
\approx
\sum_{\alpha \in \{\mathrm{Si,Ge}\}}
w_{\alpha}\,\Delta_{\alpha},
\label{eq:soc_weights}
\end{equation}
In our LCAO representation, $w_{\alpha}$ is computed from the VBM eigenvector as the normalized sum of Mulliken populations on basis functions centered on species $\alpha$; physically, it is the fraction of VBM charge localized on that species.
The species-resolved SOC shifts $\Delta_{\alpha}$ are taken from tabulated elemental spin--orbit splittings from empirical-pseudopotential parameter compilations,\cite{RiegerVogl1993} with $\Delta_{\mathrm{so}}^{\mathrm{Si}}=0.044$~eV and $\Delta_{\mathrm{so}}^{\mathrm{Ge}}=0.296$~eV.
For valence states of predominantly $p$ character, the upper ($j=3/2$-derived) branch lies $\Delta_{\mathrm{so}}/3$ above the unsplit $p$-state center, so we use $\Delta_{\alpha}\approx \Delta_{\mathrm{so}}^{\alpha}/3$ in Eq.~(\ref{eq:soc_weights}).
The SOC contribution entering Eq.~(\ref{eq:vbo}) is then
\begin{equation}
\Delta E_{v,\mathrm{SOC}}^{B/A} = \Delta E_{v}^{\mathrm{SOC}}(B) - \Delta E_{v}^{\mathrm{SOC}}(A).
\label{eq:soc_diff}
\end{equation}
This approximation is physically motivated because the SOC operator is short-ranged around nuclei and the VBM in group-IV semiconductors is predominantly $p$-like.
For representative high-Ge alloy cells ($x=0.8125$ and $0.84375$), the VBM has an average $p$-orbital weight of 88.2\% (88.0--88.3\%), supporting the predominantly $p$-like assumption used in Eq.~(\ref{eq:soc_weights}).
This construction separates the composition-dependent orbital character encoded in $w_\alpha$ from the species-resolved SOC strength $\Delta_\alpha$, enabling an efficient correction without modifying the self-consistent charge density.

Semilocal DFT typically underestimates semiconductor band gaps, which can systematically bias conduction-band edges and therefore conduction-band offsets.\cite{Heyd2003,Heyd2006,DiLiberto2021}
Screened hybrid functionals such as Heyd--Scuseria--Ernzerhof (HSE) reduce this error by replacing part of short-range semilocal exchange with exact exchange.\cite{Heyd2003,Heyd2006}
We therefore refine bulk conduction edges using HSE at selected strained Si and \SiGe\ compositions chosen to span the composition range and sample the high-$x$ regime, and combine these edges with the PBE interface-lineup term $\VIF$ from thick superlattice simulations.
We do not rerun the full 512-atom disordered superlattice simulations with HSE. Instead, we combine PBE lineup terms with HSE-refined bulk conduction edges as a practical approximation that separates the problem into interface electrostatics and functional-sensitive bulk band-edge contributions. Its use is motivated here by the isovalent Si/Ge-derived interfaces considered, for which large interface charge transfer is not expected, so that the dominant functional sensitivity is anticipated to enter through the bulk band edges more strongly than through the lineup term. \cite{VandeWalleMartin1987,Dandrea1992,DiLiberto2021}
We tested this approximation directly in the reduced 64-atom cell introduced above, computing $\VIF$ with PBE and HSE06 for identical geometries and numerical settings. The lineup changed from $-1.5843$~eV at PBE to $-1.5711$~eV at HSE06, a functional sensitivity of only $+13.2$~meV. The purely electrostatic part of the potential step is more sensitive to the functional, but the local XC contribution compensates for this in the total local potential of Eq.~(\ref{eq:vbar_bulk}), which we use consistently for bulk and interface cells. This sensitivity is comparable to the numerical checks above and small compared with the representative $\sim 30$~meV offset uncertainty, supporting the combined approach.

Figure~\ref{fig:gap_relaxed} shows the HSE band gap of relaxed \SiGe\ across composition. A clear change in slope is observed at high Ge content, in agreement with experimental data \cite{Weber1989} associated with the crossover of the conduction-band minimum from the Si-like $\Delta$ valleys to the Ge-like $L$ valleys in relaxed alloys near $x\approx 0.85$. Capturing this high-$x$ slope change provides an internal check that the bulk workflow and alloy structural modeling reproduce a key electronic-structure signature of the SiGe system relevant for conduction-band alignment.
Biaxial strain is included explicitly in all strained bulk reference cells through the imposed in-plane lattice constant and relaxed out-of-plane lattice parameter. This strain splits and shifts the conduction valleys, but a uniform coherent strain field does not by itself mix Bloch states at different crystal momenta. Alloy disorder can weakly couple states associated with different conduction-band valleys through the finite SQS potential, and this effect is included in the atomistic supercells. The CBM is selected by scanning the sampled Brillouin zone, so the high-Ge $\Delta$-to-$L$ crossover and the associated changes in the conduction-band-offset profiles enter the workflow directly.

For strained Si on \SiGe, the hybrid functional mitigates semilocal gap error that can otherwise bias conduction-band offsets through the bulk band-edge term.

\begin{figure}[t]
\centering
\includegraphics[width=\columnwidth]{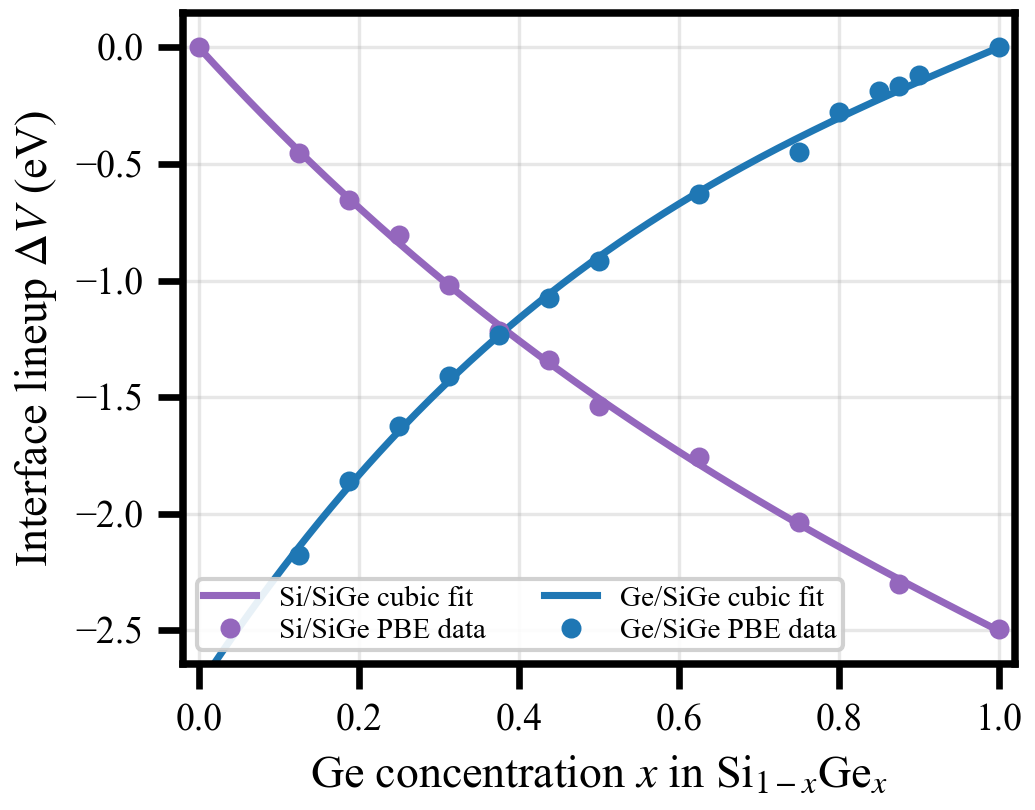}
\caption{\label{fig:lineup}
Interface lineup $\VIF$ as a function of Ge concentration $x$ in Si$_{1-x}$Ge$_x$ for strained Si/\SiGe\ and strained Ge/\SiGe, extracted from macroscopically averaged local Kohn--Sham potentials in thick periodic superlattices.
The curves are constrained cubic fits satisfying $\VIF(0)=0$ for Si/Si and $\VIF(1)=0$ for Ge/Ge.}
\end{figure}

Figure~\ref{fig:lineup} summarizes the lineup term $\VIF$ across composition for both Si/\SiGe\ and Ge/\SiGe\ interfaces.
\begin{figure*}[!t]
\centering
\includegraphics[width=0.8\textwidth]{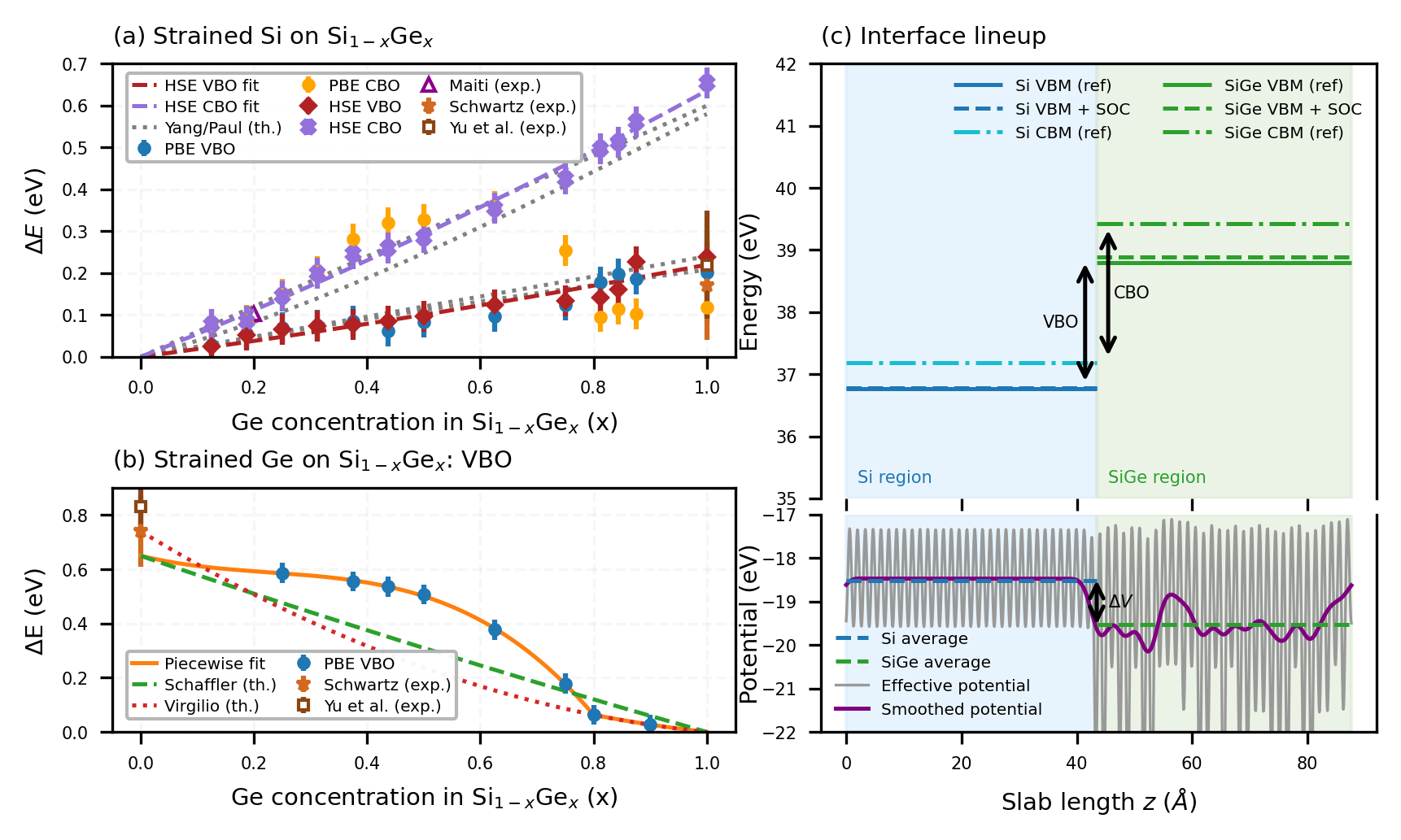}
\caption{\label{fig:offsets}
Band-offset summary.
(a) Strained-Si/\SiGe\ valence- and conduction-band offsets versus Ge concentration, with HSE fit guides and the experimental strained-Si/\SiGe\ CBO benchmark at $x=0.2$ from MOS-capacitor extraction.\cite{Maiti2004}
(b) Strained-Ge/\SiGe\ valence-band offset versus $x$, with theory curves and directly comparable endpoint photoemission data.\cite{Schaffler1997,Virgilio2006,Schwartz1989,Yu1990}
(a,b) Computed-point error bars combine two representative band-edge uncertainties inferred from the three-SQS gap spread with a 10~meV lineup term, added in quadrature; experimental bars are taken from the cited measurements when reported.
(c) Representative lineup extraction used to obtain $\VIF$.}
\end{figure*}
At compositions with repeated alloy realizations, we average over independent SQS superlattices to suppress configuration noise.
For practical use in device simulation, the interface lineup is represented by the following constrained cubic fits:
\begin{equation}
\label{eq:lineup_fits}
\begin{aligned}
\VIF^{\mathrm{Si/SiGe}}(x) &=
-0.4878\,x^{3} + 1.7497\,x^{2} - 3.7635\,x,\\
\VIF^{\mathrm{Ge/SiGe}}(x) &=
1.1837\,(x-1)^{3} - 0.1159\,(x-1)^{2} \\
&\hphantom{=}+ 1.4412\,(x-1).
\end{aligned}
\end{equation}
In Eq.~(\ref{eq:lineup_fits}), $x$ is dimensionless and all coefficients are in eV.
\vspace{-4pt}

Combining the bulk band-edge terms with the fitted lineup expressions yields the full band offsets summarized in Fig.~\ref{fig:offsets}.
For strained Si on relaxed \SiGe, HSE-refined conduction edges avoid severe semilocal-gap underestimation at large tensile strain and yield trends consistent with electron confinement in strained Si quantum wells on SiGe buffers.\cite{Schaffler1997,Paul2004,Yang2004}
For use in continuum device solvers such as QTCAD\textsuperscript{\textregistered},\cite{QTCADWebsite} the HSE-refined strained-Si offsets are
\begin{equation}
\label{eq:offset_fits}
\begin{aligned}
\DV^{\mathrm{HSE}}(x) &= 0.03848\,x^{2}+0.18158\,x,\\
\DC^{\mathrm{HSE}}(x) &= 0.10039\,x^{2}+0.53559\,x.
\end{aligned}
\end{equation}
For strained Ge on relaxed \SiGe, the corresponding piecewise fit is
\begin{equation}
\label{eq:ge_offset_fit}
\begin{aligned}
\DV^{\mathrm{Ge/SiGe}}(x) &= 0.06302 + 2.45735\,u - 4.03457\,u^{2} \\
&\hspace{1.7cm}+ 2.34954\,u^{3},\\
\DV^{\mathrm{Ge/SiGe}}(x) &= 0.22907\,v + 0.43005\,v^{2},\\
\end{aligned}
\end{equation}
Here the first line is used for $0 \le x \le 0.8$ with $u=0.8-x$, and the second for $0.8 \le x \le 1$ with $v=1-x$.
All coefficients in Eqs.~(\ref{eq:offset_fits}) and (\ref{eq:ge_offset_fit}) are in eV; Eq.~(\ref{eq:ge_offset_fit}) reproduces the sampled Ge/\SiGe\ data with a maximum absolute deviation below 7~meV.

For the lineup-derived offsets in Fig.~\ref{fig:offsets}, the explicitly sampled uncertainty sources are (i) band-edge variability across alloy realizations and (ii) sensitivity to plateau selection in the lineup extraction. These give a representative statistical uncertainty of a few $10$~meV, obtained from two 25~meV band-edge terms, inferred as half of the 50~meV three-SQS gap spread in Fig.~\ref{fig:gap_relaxed}, together with a lineup term of about 10~meV added in quadrature. In addition, unquantified systematic effects remain, including the hybrid-functional parametrization, basis and pseudopotential choices, and residual finite-size error. We therefore expect the total uncertainty in the predicted offsets to be on the order of 30~meV.

Overall, we developed a unified first-principles workflow for band alignment in strained Si/\SiGe\ and Ge/\SiGe\ heterostructures across the full composition range. Combining epitaxial lattice constants, SQS disorder modeling, potential-lineup extraction, a Mulliken-weight SOC valence correction, and hybrid-functional conduction edges, it yields smooth, internally consistent offsets and analytic expressions for continuum device simulation.

\noindent\textbf{Supplementary material.} Table~S1 gives the full band-offset dataset, lattice parameters, and literature-model comparisons.

\vspace{-10pt}
\begin{acknowledgments}
\vspace{-10pt}
This work used Digital Research Alliance of Canada computational resources. The authors thank David van Driel, Achilleas Bardakas, and L. Stehouwer for discussions of experimental band gap values.
\end{acknowledgments}

\vspace{-10pt}
\section*{Conflict of Interest}
\vspace{-10pt}
F. Beaudoin and Y. Zhu are Chief Executive Officer and President of Nanoacademic Technologies Inc., respectively. They own equity in the company.

\vspace{-10pt}
\section*{Author Contributions}
\vspace{-10pt}
\textbf{N. M. Vegh:} Methodology; investigation; formal analysis; writing---original draft. \textbf{P. Philippopoulos:} Methodology; formal analysis. \textbf{R. J. Prentki:} Methodology; writing---review and editing. \textbf{W. Zhang:} Methodology. \textbf{Y. Zhu, F. Beaudoin, and H. Guo:} Supervision.

\vspace{-10pt}
\section*{Data Availability Statement}
\vspace{-10pt}
Data, input files, and analysis scripts are available from the corresponding author upon reasonable request.

\bibliography{iopband-alignment}

\end{document}


\begin{center}
{\Large Supplementary Material}\\[0.4em]
{\normalsize for}\\[0.2em]
{\normalsize ``First-principles predictions of band alignment in strained Si/Si$_{1-x}$Ge$_x$ and Ge/Si$_{1-x}$Ge$_x$ heterostructures''}
\end{center}

\vspace{0.8em}
Supplementary Table~S1 reports the composition-resolved band-offset dataset used in this work and the benchmark columns discussed in the main text.

\begin{table}[h!]
\caption{Master band-offset dataset for strained Si/Si$_{1-x}$Ge$_x$ and Ge/Si$_{1-x}$Ge$_x$ across composition.
$a_{\parallel}$ is the in-plane lattice constant (biaxial strain setting) in \AA.
Offsets are in eV. For strained Si/SiGe, we report this work at the PBE level and with HSE-refined bulk band edges (lineup from PBE), alongside Paul benchmarks. For strained Ge/SiGe, we report the PBE valence-band offset and include the Virgilio benchmark. Entries shown as $\cdots$ are not available in the present dataset.}
\centering
\scriptsize
\setlength{\tabcolsep}{3.0pt}
\renewcommand{\arraystretch}{1.15}
\resizebox{\textwidth}{!}{%
\begin{tabular}{c c cc cc cc cc}
\toprule
& & \multicolumn{6}{c}{Strained Si on relaxed Si$_{1-x}$Ge$_x$} & \multicolumn{2}{c}{Strained Ge on relaxed Si$_{1-x}$Ge$_x$} \\
\cmidrule(lr){3-8}\cmidrule(lr){9-10}
$x$ & $a_{\parallel}$ (\AA)
& \multicolumn{2}{c}{This work (PBE)}
& \multicolumn{2}{c}{This work (HSE-refined)}
& \multicolumn{2}{c}{Paul (benchmark)}
& \multicolumn{1}{c}{This work (PBE)}
& \multicolumn{1}{c}{Virgilio (benchmark)} \\
\cmidrule(lr){3-8}\cmidrule(lr){9-10}
& & $\Delta E_v$ (VBO) & $\Delta E_c$ (CBO)
  & $\Delta E_v$ (VBO) & $\Delta E_c$ (CBO)
  & $\Delta E_v$ (VBO) & $\Delta E_c$ (CBO)
  & $\Delta E_v$ (VBO)
  & $\Delta E_v$ (VBO) \\
\midrule
0.1250 & 5.496 & 0.0288 & 0.0763 & 0.0257 & 0.0776 & 0.0300 & 0.0750 & \multicolumn{1}{c}{$\cdots$} & \multicolumn{1}{c}{$\cdots$} \\
0.1875 & 5.511 & 0.0555 & 0.0870 & 0.0513 & 0.0857 & 0.0450 & 0.1125 & \multicolumn{1}{c}{$\cdots$} & \multicolumn{1}{c}{$\cdots$} \\
0.2500 & 5.524 & 0.0698 & 0.1498 & 0.0661 & 0.1446 & 0.0600 & 0.1500 & 0.5861 & 0.4556 \\
0.3125 & 5.540 & 0.0750 & 0.2030 & 0.0736 & 0.2000 & 0.0750 & 0.1875 & \multicolumn{1}{c}{$\cdots$} & \multicolumn{1}{c}{$\cdots$} \\
0.3750 & 5.556 & 0.0851 & 0.2818 & 0.0769 & 0.2463 & 0.0900 & 0.2250 & 0.5545 & 0.3383 \\
0.4375 & 5.575 & 0.0609 & 0.3198 & 0.0856 & 0.2599 & 0.1050 & 0.2625 & 0.5359 & 0.2858 \\
0.5000 & 5.595 & 0.0831 & 0.3285 & 0.0974 & 0.2842 & 0.1200 & 0.3000 & 0.5073 & 0.2375 \\
0.6250 & 5.642 & 0.0973 & 0.3582 & 0.1246 & 0.3557 & 0.1500 & 0.3750 & 0.3767 & 0.1533 \\
0.7500 & 5.696 & 0.1236 & 0.2543 & 0.1330 & 0.4256 & 0.1800 & 0.4500 & 0.1786 & 0.0856 \\
0.8000 & 5.714 & \multicolumn{1}{c}{$\cdots$} & \multicolumn{1}{c}{$\cdots$} & \multicolumn{1}{c}{$\cdots$} & \multicolumn{1}{c}{$\cdots$} & \multicolumn{1}{c}{$\cdots$} & \multicolumn{1}{c}{$\cdots$} & 0.0630 & 0.0632 \\
0.8125 & 5.722 & 0.1786 & 0.0957 & 0.1420 & 0.4968 & 0.1950 & 0.4875 & \multicolumn{1}{c}{$\cdots$} & \multicolumn{1}{c}{$\cdots$} \\
0.8438 & 5.736 & 0.1973 & 0.1144 & 0.1605 & 0.5130 & 0.2025 & 0.5062 & \multicolumn{1}{c}{$\cdots$} & \multicolumn{1}{c}{$\cdots$} \\
0.8750 & 5.751 & 0.1860 & 0.1031 & 0.2272 & 0.5617 & 0.2100 & 0.5250 & \multicolumn{1}{c}{$\cdots$} & \multicolumn{1}{c}{$\cdots$} \\
0.9000 & 5.756 & \multicolumn{1}{c}{$\cdots$} & \multicolumn{1}{c}{$\cdots$} & \multicolumn{1}{c}{$\cdots$} & \multicolumn{1}{c}{$\cdots$} & \multicolumn{1}{c}{$\cdots$} & \multicolumn{1}{c}{$\cdots$} & 0.0272 & 0.0263 \\
1.0000 & 5.802 & 0.2010 & 0.1181 & 0.2380 & 0.6540 & 0.2400 & 0.6000 & 0.0000 & 0.0000 \\
\bottomrule
\end{tabular}%
}
\end{table}